\documentclass[mathpazo]{cicp}
\usepackage{subcaption}
\usepackage{graphicx}
\usepackage{mathtools}

\begin{document}

\title{Analytical and Computational Studies of Correlations of Hydrodynamic Fluctuations in Shear Flow}
\author[Bian, Deng and Karniadakis]{Xin Bian\corrauth \footnote[2]{present address: Lehrstuhl f\"ur Aerodynamik und Str\"omungsmechanik, Technische Universit\"at M\"unchen, 85748 Garching, Germany}, 
Mingge Deng, and George Em Karniadakis}
\address{Division of Applied Mathematics, Brown University, Providence, Rhode Island 02912, USA}
% \emails{{\tt xin.bian@tum.de} (Xin Bian)
% {\tt george\_karniadakis@brown.edu} (George Em Karniadakis)}
\email{{\tt xin.bian@tum.de} (Xin Bian)}

\begin{abstract}
We study correlations of hydrodynamic fluctuations in shear flow
analytically and also by dissipative particle dynamics~(DPD) simulations.
The hydrodynamic equations are linearized around the macroscopic velocity field
and then solved by a perturbation method in Fourier-transformed space.
The autocorrelation functions~(ACFs) from the analytical method are compared with 
results obtained from DPD simulations
under the same shear-flow conditions.
Upto a moderate shear rate, 
various ACFs from the two approaches agree with each other well.
At large shear rates, discrepancies between the two methods are observed,
hence revealing strong additional coupling between different fluctuating variables,
which is not considered in the analytical approach.
In addition, the results at low and moderate shear rates can serve as benchmarks for 
developing multiscale algorithms for coupling of heterogeneous solvers,
such as a hybrid simulation of molecular dynamics and fluctuating hydrodynamics solver,
where thermal fluctuations are indispensable.
\end{abstract}

\maketitle

\section{Introduction}
The complex behavior of many particles plays a significant role
in atomic fluids~\cite{Hansen2013, Karniadakis2005},
chemical and biological processes~\cite{Mewis2012, Li2014b},
granular materials~\cite{Jaeger1996}, and astrophysics~\cite{Springel2010}.
On the one hand, given interparticle potentials,
a kinetic-theory type of description from first principles can be formulated,
which, however, may be too complex to apply.
On the other hand, 
if physical quantities vary rather slowly in space and time,
a local thermodynamic equilibrium may be valid.
Therefore, the system can be represented by continuous hydrodynamic fields
described by the Navier-Stokes-Fourier (NSF) equations,
which take into account the conservation of mass, momentum and energy.
Although such partial differential equations (PDEs) are concise and practically powerful,
the thermodynamic derivatives and transport coefficients
must be obtained from a more fundamental theory to complete the phenomenological description.
Through the seminal efforts of Einstein, Onsager, Callen, Welton, Green, Kubo
and many others~\cite{Kubo1966, Hansen2013},
a general linear response theory has been established.
Furthermore, the transport coefficients are connected to the
corresponding correlation functions~(CFs) of the microscopic fluctuating variables.
These connections are all embraced in the fundamental Green-Kubo relations.
At equilibrium or small deviations from equilibrium, 
a systematic connection between the CFs and
the hydrodynamics equations has been established
for the long wave-length and small-frequency hydrodynamic limit~\cite{Kadanoff1963, Boon1991, Hansen2013}.
In this hydrodynamic limit, the effects of the solvent fluctuations
on suspended Brownian particles have also been studied extensively~\cite{Bian2016a}.
A further extension for small wave-length of the fluid has been made,
which connects the microscopic dynamics to the generalized hydrodynamic equations~\cite{Boon1991}.
Another breakthrough is the development of the fundamental fluctuation relations
at transient or stationary nonequilibrium state far from equilibrium,
which was initiated by Evans et al.~\cite{Evans1993}
and has later engaged many others~\cite{Evans2008, Seifert2012, Bertini2015}.
Many of these works on statistical mechanics are closely related to the large deviation theory in probability theory~\cite{Marconi2008}.

At a stationary nonequilibrium state,
it seems relatively simpler to start with the phenomenological hydrodynamic equations and work reversely
to obtain various CFs of the fluctuating variables~\cite{Machta1980, Tremblay1981, Garcia1987a, MalekMansour1987, Farrell1993, Wada2004, OrtizdeZarate2008, OrtizdeZarate2009, Zhang2009a, OrtizdeZarate2011}.
This strategy has been applied by Lutsko\&Dufty~\cite{Lutsko1985}
and has been receiving continuous attention~\cite{Otsuki2009, Otsuki2009a, Varghese2015}.
In the present work, we consider an isothermal shear flow at steady state 
as a typical setting for the nonequilibrium behavior of many particles.
Following the pioneering derivations of Lutsko\&Dufty~\cite{Lutsko1985}, 
the equations of fluctuating hydrodynamics are linearized around the steady state by assuming small fluctuations,
before they are transformed into the Fourier space.
Thereafter, an equivalent generalized eigenvalue problem is solved perturbatively
to provide the temporal evolution of the hydrodynamic fluctuations.
Finally, various autocorrelation functions~(ACFs) can be constructed in the Fourier space
and transformed into the real space if needed.
Some analytical ACFs have been recently compared with 
inelastic hard-sphere simulations and multi-particle collision dynamics 
at low shear rates~\cite{Otsuki2009, Otsuki2009a, Varghese2015}.
As the first objective of this work, we aim to verify the analytical ACFs at low/moderate shear rates
and search for deviations from the theory at large shear rates via numerical simulations.
We expect that our computations will reveal a transition
from decoupling to the coupling of different fluctuating variables and
may shed light on the possible extension of the theory at large shear rates.
To this end, we employ a mesoscopic method called dissipative particle dynamics~(DPD)
to quantify such deviations.

The DPD method describes the behavior of many particles at mesoscale and
was invented to bridge the gap between the microscopic dynamics and the macroscopic PDEs~\cite{Hoogerbrugge1992}.
In a DPD system, three pairwise-additive forces 
${\bf F}^C_{ij}$, ${\bf F}^D_{ij}$ and ${\bf F}^R_{ij}$ are prescribed
between neighboring particles $i$ and $j$ and they correspond to the 
underlying conservative, dissipative and random process, respectively~\cite{Espanol1995, Groot1997}.
By postulating a steady state solution of the corresponding Fokker-Planck equations	,
${\bf F}^D_{ij}$ and ${\bf F}^R_{ij}$ are found to correlate with each other
so that the fluctuation-dissipation theorem is satisfied
and the canonical ensemble is warranted~\cite{Espanol1995}.
Given molecular dynamics (MD) trajectories,
the pairwise forces in DPD may be constructed via a coarse-graining procedure
following the Mori-Zwanzig formalism~\cite{Hijon2010, Lei2010, Li2014a}.
Without data from a reference MD simulation, the parameters of the pairwise forces are usually tuned
to achieve static and dynamic properties of a target fluid empirically~\cite{Fan2003, Li2016}.
Although the kinetic theory for the DPD particles
can qualitatively predict the transport coefficients~\cite{Marsh1997a, Groot1997},
a quantitative knowledge is only available via {\it a posteriori} processing of the simulation results~\cite{Groot1997, Fan2003, Backer2005, Fedosov2010a, Lei2011, Bian2015a, Azarnykh2016}.
DPD typically has a softer potential between particles than that of MD,
therefore it allows for a larger time step.
In the hydrodynamic limit with large spatial-temporal scales,
it may be considered as a Lagrangian discrete counterpart of the fluctuating hydrodynamics
described by the Landau-Lifshitz-Navier-Stokes equations~\cite{Landau1959, Bian2015a}.
At small wave-length and high frequency, 
it may be considered as the representation of the generalized hydrodynamics~\cite{Ripoll2001, Azarnykh2016},
especially when the pairwise forces of DPD are obtained via the coarse-graining
and non-Markovian effects are nonnegligible~\cite{Li2015c, Lei2016b, Li2017}.
In addition to simulations of simple fluid at mesoscale,
DPD also finds wide applications in simulating complex fluids such as
colloidal suspension, polymer solution and red blood cells under flow~\cite{Pan2009, Bian2012, Groot1997, Fan2003, Li2014b}.
As the second goal of this work,
for the first time we evaluate various ACFs generated
by DPD simulations under shear flow by comparing with analytical solutions.
This would provide a solid evidence as to when DPD may be an effective solver
of the fluctuating hydrodynamics at nonequilibrium.

Recently, multiscale coupling of heterogeneous solvers (e.g., molecular dynamics and Navier-Stokes solver)
has been attracting a lot of attention~\cite{E2007, Praprotnik2008, Delgado-Buscalioni2012, Borg2015, Bian2015, Tang2015, Perdikaris2015}.
Both the accurate dynamics of a fine model and the computational efficiency of a coarse model
my be exploited in such a hybrid simulation.
In the course of the coupling,
from a continuum perspective the thermal fluctuations are treated very often as unwanted noises to be filtered out.
However, the fluctuations are unique hallmarks to micro-/meso-scopic physics within a finite volume of material
and their space-time correlations encode the full dynamic information of the system~\cite{Hansen2013}.
Therefore, as the third goal of this work,
we advocate that the various ACFs should be taken as benchmarks
for a hybrid simulation whenever it is possible.
For example, one transversal ACF was previously compared among different coupling algorithms for a shear flow,
which led to uncovering certain artifacts of the specific coupling~\cite{Bian2016}.

We shall proceed as follows.
In Section~\ref{section_theory} we shall revisit the linearized fluctuating hydrodynamics
and construct the temporal ACFs of the fluctuating variables in k-space.
In Section~\ref{section_dpd} we will describe the DPD method with Lees-Edwards boundary conditions,
and further elaborate on some technical details on the implementation of DPD simulations.
In Section~\ref{section_results}, we compare the theory with simulations 
for two set of transversal autocorrelation functions~(TACFs) and
longitudinal autocorrelation functions~(LACFs) for various wave vectors
for a range of shear rates.
Finally, we summarize our findings in Section~\ref{section_summary} with discussions.
Extra details on the theoretical derivations are given in Appendices~\ref{appendix_linearization} and \ref{appendix_hydrodynamic_matrix_and_modes}.

\section{Theory revisited}
\label{section_theory}

In this section, we follow the pioneering work of Lutsko\&Dufty~\cite{Lutsko1985, Lutsko1986a}
to derive analytically the ACFs of fluctuating variables in shear flow.
Some of the calculations have also recently been performed~\cite{Otsuki2009, Otsuki2009a, Varghese2015}.
Firstly, we describe the equations of fluctuating hydrodynamics in Section~{\ref{section_theory_hydrodynamics}.
Subsequently, in Appendix~\ref{appendix_linearization} we linearize the equations around the macroscopic or averaged state
by keeping only the first-order fluctuating variables.
Thereafter, in Appendix~\ref{appendix_hydrodynamic_matrix_and_modes} we perform a spatial Fourier transform of the linearized equations.
By applying a perturbation theory,
we find the approximate solutions by solving a generalized eigenvalue problem.
Finally, we summarize various ACFs of the fluctuating variables in Section~{\ref{section_theory_correlations}.

\subsection{Fluctuating hydrodynamics}
\label{section_theory_hydrodynamics}

Due to the conservation of mass and momentum,
the equations of continuity and dynamics for an isothermal fluid read as, 
\begin{eqnarray}
 \left( \frac{\partial}{\partial t} + {\bf v} \cdot \nabla \right) \rho &=& - \rho \nabla \cdot {\bf v}, \label{eq_continuity_Eulerian} \\
 \rho \left( \frac{\partial}{\partial t} + {\bf v} \cdot \nabla \right) {\bf v} &=& \nabla \cdot {\bf \Pi}, \label{eq_momentum_Eulerian} 
\end{eqnarray}
where $\rho$ is the mass density, ${\bf v}$ is the velocity
and ${\bf \Pi}$ is the stress tensor.
By defining the particle derivative as
$\frac{d}{dt} = (\frac{\partial}{\partial t} + {\bf v} \cdot \nabla)$,
we have the hydrodynamic equations in the Lagrangian form as
\begin{eqnarray}
 \frac{d \rho}{dt} &=& - \rho \nabla \cdot {\bf v} , \label{eq_continuity_Lagrangian} \\
 \rho\frac{d {\bf v}}{dt} &=& \nabla \cdot {\bf \Pi}. \label{eq_momentum_Lagrangian} 
\end{eqnarray}
The Lagrangian form of the hydrodynamic equations
may be particularly useful to interpret the DPD method described in Section \ref{section_dpd}.

The stress tensor $\Pi_{\mu\sigma}$ may be considered as a linear combination of 
three components
\begin{eqnarray}
\Pi_{\mu\sigma}   &=& \Pi^C_{\mu\sigma} + \Pi^D_{\mu\sigma} + \Pi^R_{\mu\sigma}.
\end{eqnarray}
Each of the components is according to the reversible, irreversible, and stochastic process, respectively.
Assuming that the pressure is isotropic and the viscous stress depends only on the first derivatives of velocity,
$\Pi^C$ and $\Pi^D$ read as~\cite{Landau1959}
\begin{eqnarray}
\Pi^C_{\mu\sigma} &=& -p  \delta_{\mu\sigma},\\
\Pi^D_{\mu\sigma} &=& \eta \left(\frac{\partial v_{\mu}}{\partial x_{\sigma}} + \frac{\partial v_{\sigma}}{\partial x_{\mu}} 
      -\frac{2}{3} \delta_{\mu\sigma} \frac{\partial v_{\epsilon}}{\partial v_{\epsilon}} \right) 
      + \zeta \delta_{\mu\sigma}\frac{\partial v_{\epsilon}}{\partial x_{\epsilon}},
\end{eqnarray}
where $\eta$ and $\zeta$ are constant coefficients of the shear and bulk viscosity.
Summation convention for Greek indices is adopted and
$\delta_{\mu\sigma}$ is the Kronecker $\delta$.
Furthermore, 
the pressure $p=p(\rho)$ is determined by an equation of state at equilibrium.
This completes the description for the compressible Navier-Stokes equations
\begin{eqnarray}
 \rho \left( \frac{\partial}{\partial t} + {\bf v} \cdot \nabla \right) {\bf v} = 
 -\nabla p + \eta \nabla^2 {\bf v} + \left( \frac{\eta}{3}+\zeta \right) \nabla \left( \nabla \cdot {\bf v} \right). \label{eq_momentum_Navier-Stokes} 
\end{eqnarray}

For fluids at mesoscale, there are fluctuations in the state variables governed by the framework of thermodynamics,
therefore local spontaneous stress does occur.
By assuming an underlying Gaussian-Makovian process for the unresolved degrees of freedom,
the conditions for the random stress tensor satisfying the fluctuation-dissipation theorem read as~\cite{Landau1959}
\begin{eqnarray}
 <\Pi^R_{\mu\sigma}> &=& 0, \\
 <\Pi^R_{\mu\sigma}({\bf x}, t)\Pi^R_{\epsilon \iota}({\bf x}', t')> &=& 2k_BT \Delta_{\mu\sigma\epsilon l}\delta({\bf x}-{\bf x}')\delta(t-t'), \\
 \Delta_{\mu\sigma\epsilon \iota} &=& 
 \eta \left(\delta_{\mu \epsilon}\delta_{\sigma \iota}+ 
 \delta_{\mu \iota}\delta_{\sigma\epsilon} \right) +\left( \zeta - \frac{2}{3}\eta \right) \delta_{\mu\sigma} \delta_{\epsilon \iota}.
\end{eqnarray}
Here $\delta({\bf x}-{\bf x}')$ or $\delta(t-t')$ is the Dirac $\delta$ function.
This completes the description for the Landau-Lifshitz-Navier-Stokes equations,
which sometimes are also referred to as the Navier-Stokes-Langevin equations~\cite{Lutsko1985}.

\subsection{Autocorrelation functions under shear flow}
\label{section_theory_correlations}
For a uniform shear flow, the macroscopic stationary state of the velocity field ${\bf v}_0$ reads as
\begin{eqnarray}
 v_{0\mu}({\bf x}, t) &=& \dot \gamma_{\mu\sigma} x_{\sigma}, \label{eq_shear_flow1}\\
\dot \gamma_{\mu\sigma} &=& \dot \gamma \delta_{\mu x}\delta_{\sigma y}. \label{eq_shear_flow2}
\end{eqnarray}
This corresponds to a flow along the $x$ direction,
with velocity gradient $\dot \gamma$ in the $y$ direction,
and vorticity along the $z$ direction.
By assuming {\it small deviations} from the averaged fields,
the fluctuating hydrodynamic equations of Eqs.~(\ref{eq_continuity_Eulerian}) and (\ref{eq_momentum_Eulerian})
may be linearized as shown in Appendix~\ref{appendix_linearization},
where second-order terms in fluctuations are dropped off.
Afterwards, a system of linear equations may be spatial-Fourier transformed
to obtain the hydrodynamic equations (\ref{eq_hydrodynamic_equation}) in k-space.
The general solution to Eq.~(\ref{eq_hydrodynamic_equation})
is determined from a generalized eigenvalue problem of Eq.~(\ref{eq_eigenvalue_equation}).
The latter is solved via a perturbation theory by
expanding the size of wave vector $k=|{\bf k}|$ as a small parameter to second order in the continuum limit~\cite{Lutsko1985, Lutsko1986a}.
Technical details of the derivations are elaborated in Appendices~\ref{appendix_linearization} and \ref{appendix_hydrodynamic_matrix_and_modes}.

Here, we need to emphasize one important assumption of the perturbation method.
The shear rate is assumed to be {\it moderate} so that $\dot \gamma \lesssim \nu k^2 \ll c_Tk$.
Hence, the term of $\dot \gamma$ is treated as in the order of $k^2$ during the perturbation.
With the kinematic viscosity $\nu$ fixed,
we may expect the perturbation method to fail
for a very large $\dot \gamma$ or very small $k$.
For our simulations in a cubic box with length $L$ in Section~\ref{section_results}, 
the size of wave vector is $|{\bf k}|=|2\pi(n_x, n_y, n_z)/L| $,
where $n_i$ is an integer and $ n^2_x+n^2_y+n^2_z \ge 1$.
If a proper box length $L$ is selected,
there is a minimal infrared cut off as $|{\bf k}|=2\pi/L$.
Therefore, the shear rate $\dot \gamma$ is the only free parameter
in the DPD simulations that we perform to validate the theory.

Given the derivations in Appendix~\ref{appendix_hydrodynamic_matrix_and_modes}, 
the longitudinal and two transversal ACFs are readily constructed as follows~\cite{Lutsko1985, Otsuki2009a, Varghese2015}
\begin{eqnarray}
\label{eq_acfs}
C_L({\bf k}, \tau) &=& \frac{<\widehat{\delta u}_1({\bf k}, t+\tau)\widehat{\delta u}_1(-{\bf k}, t)>}{<\widehat{\delta u}_1({\bf k}, t)\widehat{\delta u}_1(-{\bf k}, t)>}
		  = \left(\frac{k(\tau)}{k_0}\right)^{1/2} e^{-\Gamma_T \alpha({\bf k}, \tau)} cos(c_T\beta({\bf k}, \tau)), \label{eq_lacf} \\
C_{T_1}({\bf k}, \tau) &=& \frac{ <\widehat{\delta u}_2({\bf k}, t+\tau)\widehat{\delta u}_2 (-{\bf k}, t) >}{<\widehat{\delta u}_2({\bf k}, t)\widehat{\delta u}_2(-{\bf k}, t)>} 
                    = \left(\frac{k_0}{k(\tau)}\right)e^{-\nu \alpha({\bf k}, \tau)}, \label{eq_tacf1} \\
C_{T_2}({\bf k}, \tau) &=& \frac{<\widehat{\delta u}_3({\bf k}, t+\tau)\widehat{\delta u}_3(-{\bf k}, t) >}{<\widehat{\delta u}_3({\bf k}, t)\widehat{\delta u}_3(-{\bf k}, t)>} 
                  = e^{-\nu \alpha({\bf k}, \tau)}, \label{eq_tacf2}
\end{eqnarray}
where kinematic viscosity $\nu=\eta/\rho_0$, sound speed and attenuation coefficient are $c_T$ and $\Gamma_T=(2\eta/3+\zeta/2)/\rho_0$.
Here $\widehat {\delta u}_1({\bf k},t)$ is the longitudinal component along wave vector,
while $\widehat {\delta u}_2({\bf k},t)$ and $\widehat {\delta u}_2({\bf k},t)$ are two transversal components,
as defined in Eq.~(\ref{eq_velocity_fourier_component}).
Therefore, $C_L({\bf k},\tau)$ is the normalized longitudinal autocorrelation function (LACF),
while $C_{T_1}({\bf k},\tau)$ and $C_{T_2}({\bf k},\tau)$ are the normalized first and second
transversal autocorrelation functions (TACFs), respectively.
As a matter of fact, the normalized ACFs are just the corresponding propagator of Eq.~(\ref{eq_propagator}).
Moreover, $\alpha$ and $\beta$ are defined as 
\begin{eqnarray}
 \alpha({\bf k}, t)&=& k^2_0t-\dot \gamma k_x k_y t^2 + \frac{1}{3}\dot \gamma^2 k^2_x t^3, \label{eq_alpha}\\
 \beta({\bf k}, t) &=& \frac{1}{2\dot \gamma k_x} \left \{  k_yk_0-k_y(t)k(t)
-k^2_{\bot}  \ln  \left[ \frac{k_y(t)+k(t)}{k_y+k_0} \right]  \right \}, \label{eq_beta}
\end{eqnarray}
where the wave vector is {\it time dependent} as ${\bf k}(t)=(k_x, k_y-\dot\gamma t k_x, k_z)$ to account for the advection.
It is simple to see that when $\dot \gamma=0$,
${\bf k}(t) \equiv {\bf k}(0) = {\bf k}_0=(k_x,k_y,k_z$) and $\alpha({\bf k}, t)=k^2t$.
Therefore, Eqs. (\ref{eq_tacf1}) and (\ref{eq_tacf2})
are identical and degenerate to the solutions for equilibrium~\cite{Boon1991, Hansen2013, Bian2015a}.

To have a better sense of how the shear flow alters the dissipation and frequency of sound propagation,
we plot the functions $\alpha({\bf k}, t)$ and $\beta({\bf k}, t)$ for three representative wave vectors
in Fig. \ref{fig_exponent} and \ref{fig_frequency}, respectively.
For each wave vector ${\bf k}$, 
we also represent the equilibrium behavior $\dot\gamma=0$ with solid line as a reference accordingly.
For example, in Fig. \ref{fig_exponent}(b) with both $k_x>0$ and $k_y>0$,
the terms with $t^2$ and $t^3$ compete with each other as time progresses,
which is indicated by Eq. (\ref{eq_alpha}).
As a consequence, the dissipation rate of $\dot \gamma >0$ is slower than that of $\dot \gamma =0$ 
at short time while the reverse is true at long time.
The same phenomena in Figs. \ref{fig_exponent}(a) and (c) are relatively simpler,
hence, indicating a monotonic increase of dissipation rate with shear rate $\dot \gamma$.

\begin{figure}
\centering
\begin{subfigure}[b]{0.328\columnwidth}
  \includegraphics[width=\columnwidth, height=0.9\columnwidth]{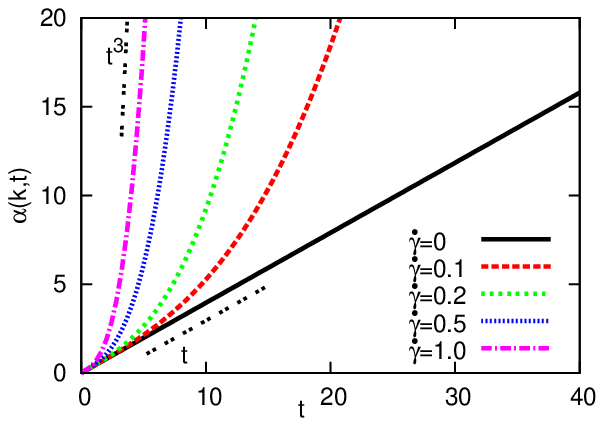}
  \caption{${\bf k}=(2\pi/L,0,0)$}
\end{subfigure}
\begin{subfigure}[b]{0.328\columnwidth}
  \includegraphics[width=\columnwidth, height=0.9\columnwidth]{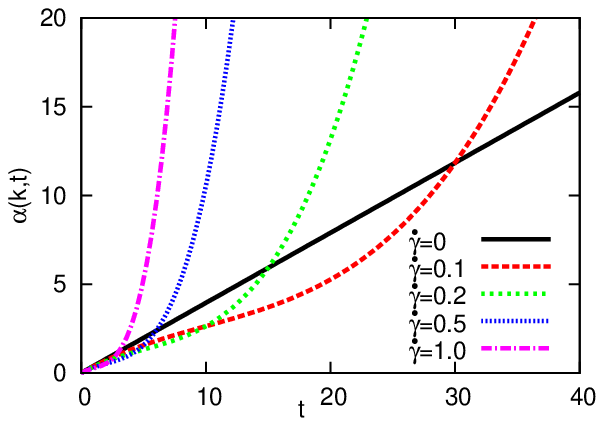}
  \caption{${\bf k}=(\sqrt{2}\pi/L,\sqrt{2}\pi/L,0)$}
\end{subfigure}
\begin{subfigure}[b]{0.328\columnwidth}
  \includegraphics[width=\columnwidth, height=0.9\columnwidth]{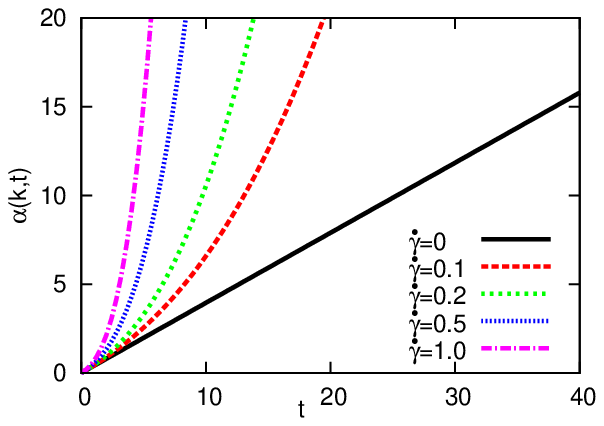}
  \caption{${\bf k}=(\sqrt{2}\pi/L,-\sqrt{2}\pi/L,0)$}
\end{subfigure}
  \caption{Exponent of dissipation in Eq.~(\ref{eq_alpha}) for different wave vectors in shear flow: 
  a typical wave length $L=10$ is taken.}
  \label{fig_exponent}
 \end{figure}

Similarly, from Eq. (\ref{eq_beta}) we may understand how the shear flow affects the frequency of sound propagation.
For example, in Fig. \ref{fig_frequency}(b) with $k_x>0$ and $k_y>0$,
the constant slope of $\beta({\bf k},t)$ at $\dot \gamma=0$ indicates a constant frequency at equilibrium,
whereas $\dot \gamma>0$ corresponds to frequency decrease at short time and frequency increase at long time.
The same phenomena in Figs. \ref{fig_frequency}(a) and (c) are less complicated,
that is, the shear rate always enhances the sound frequency.
From Eq. (\ref{eq_beta}) and Fig. \ref{fig_frequency}, 
we may expect Doppler effects for the LACFs.

\begin{figure}
\centering
\begin{subfigure}[b]{0.328\columnwidth}
  \includegraphics[width=\columnwidth, height=0.9\columnwidth]{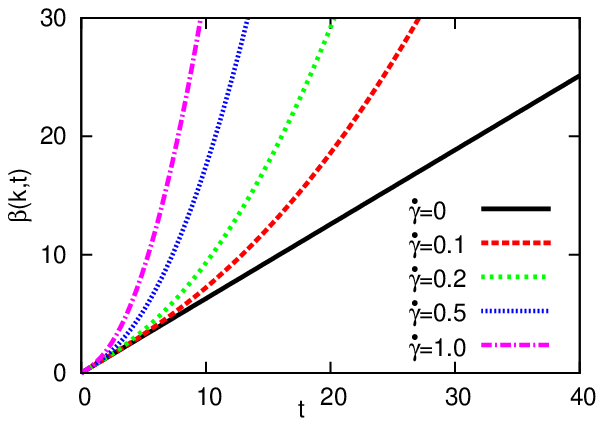}
  \caption{${\bf k}=(2\pi/L,0,0)$}
\end{subfigure}
\begin{subfigure}[b]{0.328\columnwidth}
  \includegraphics[width=\columnwidth, height=0.9\columnwidth]{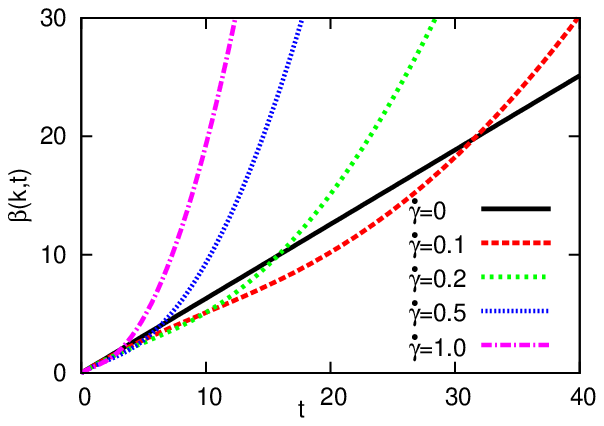}
  \caption{${\bf k}=(\sqrt{2}\pi/L,\sqrt{2}\pi/L,0)$}
\end{subfigure}
\begin{subfigure}[b]{0.328\columnwidth}
  \includegraphics[width=\columnwidth, height=0.9\columnwidth]{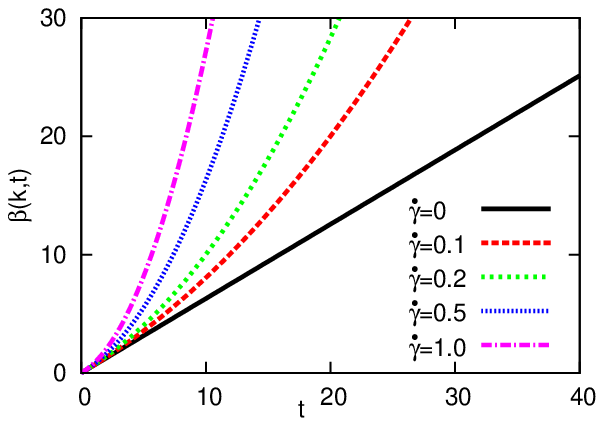}
  \caption{${\bf k}=(\sqrt{2}\pi/L,-\sqrt{2}\pi/L,0)$}
\end{subfigure}
  \caption{Frequency of sound propagation in Eq.~(\ref{eq_beta}) for different wave vectors in shear flow: 
  a typical wave length $L=10$ is taken.}
  \label{fig_frequency}
 \end{figure}

\section{Dissipative particle dynamics}
\label{section_dpd}
The method of dissipative particle dynamics (DPD) was invented two decades ago 
to simulate rheological properties of complex fluids at mesoscale~\cite{Hoogerbrugge1992}.
In this section, we shall briefly revisit the classical DPD method
and its implementation of boundary conditions in shear flow.
\subsection{Pairwise forces}
\label{section_dpd_force}
For convenience,  as reference we define some simple notations
\begin{eqnarray}
 {\bf r}_{ij}   &=& {\bf r}_i-{\bf r}_j, \nonumber \\
 {\bf v}_{ij}   &=& {\bf v}_i-{\bf v}_j, \nonumber \\
 {\bf e}_{ij}   &=& {\bf r}_{ij}/ r_{ij}, \quad
 r_{ij}         = |{\bf r}_{ij}|,    
 \label{eq_notation_rvg}
\end{eqnarray}
where ${\bf r}_i$, ${\bf v}_i$ are position and velocity of particle $i$;
${\bf r}_{ij}$, ${\bf v}_{ij}$ are relative position and velocity of particles $i$ and $j$;
$r_{ij}$ is the distance between the two and 
${\bf e}_{ij}$ is the unit vector pointing $j$ to $i$.
The three pairwise forces are described as follows~\cite{Espanol1995, Groot1997},
\begin{eqnarray}
{\bf F}^C_{ij} &=& a W^C(r_{ij}) {\bf e}_{ij}, \\
{\bf F}^D_{ij}  &=& -\gamma W^D(r_{ij}) ({\bf e}_{ij} \cdot {\bf v}_{ij}){\bf e}_{ij}, \\
{\bf F}^R_{ij}  &=& \sigma W^R(r_{ij}) \theta_{ij} {\bf e}_{ij} \delta t^{-1/2},
\label{eq_dpd_EoM}
\end{eqnarray}
where coefficients $a$, $\gamma$, and $\sigma$ reflect the strength of individual forces;
$W^C$, $W^D$, and $W^R$ are weighting functions that monotonically decay with the relative distance 
$r_{ij}$;
$\theta_{ij}=\theta_{ji}$ is a Gaussian white noise with
\begin{eqnarray}
 <\theta_{ij}(t)> &=& 0,  \\
<\theta_{ij}(t)\theta_{kl}(t')> &=& \left(\delta_{ik}\delta_{jl} + 
\delta_{il}\delta_{jk}\right)\delta(t-t').
\end{eqnarray}
%where $\delta_{ij}$ is the Kronecker $\delta$ and $\delta(t-t')$ is the Dirac $\delta$ function.
The DPD version of the fluctuation-dissipation theorem reads as
\begin{eqnarray}
  W^D(r_{ij}) &=& \left[ W^R(r_{ij}) \right] ^ 2,\\
 2k_BT \gamma &=& \sigma^2,
\label{eq_fluctuation_dissipative_theorem}
\end{eqnarray}
which warrants the canonical ensemble~\cite{Espanol1995}.

Given the underlying force fields of molecular dynamics (MD),
the actual forms of the three pairwise forces
may be constructed via the Mori-Zwanzig projection~\cite{Hijon2010, Li2014a}.
Without referring to any particular MD system,
a typical empirical form of the weighting kernel is suggested as~\cite{Groot1997, Fan2003}
\begin{equation}
 W^{C, R}(r_{ij}) = \left \{
\begin{array}{l r}
 (1 - r_{ij}/r_c)^k, & r_{ij} < r_c, \\
 0, & r_{ij} \geq r_c.
 \end{array}
\right.
\label{eq_weighting_r}
\end{equation}
Following~\cite{Bian2016},
we take $a=25.0$, $\sigma=3.0$, $\gamma=4.5$, $r_c=1$ and $k_BT=1.0$;
$k=1$ for $W^C$ and $k=0.25$ for $W^R$;
particle mass $m=1$, number density $n=3.0$
and mass density $\rho=mn=3.0$.
For this particular set of input parameters,
the dynamic viscosity, bulk viscosity and isothermal sound speed of the fluid
are $\eta=1.62$, $\zeta=2.3$ and $c_T=4.05$ in DPD units~\cite{Bian2015a}.
The velocity Verlet time integrator is employed~\cite{Groot1997}
and $\delta t=0.005$ is small enough for stability.

\subsection{Boundary conditions}
\label{section_dpd_bc}
At nonequilibrium, the equal-time correlations of fluctuations are typically long-ranged~\cite{Otsuki2009} 
and we do not wish to introduce any extra complexities due to the boundary effects from the wall~\cite{OrtizdeZarate2013}.
Therefore, we focus on the bulk behavior of the fluid in a periodic system,
which also corresponds to the condition of the theory in Section~\ref{section_theory}.
Suppose there is a simple shear flow defined as in Eq.~(\ref{eq_shear_flow1}),
that is,  the flow is in the $x$ direction, 
the velocity gradient $\dot\gamma$ is along the $y$ direction,
and the vorticity is along the $z$ direction.
The usual periodic boundary conditions apply in the $x$ and $z$ directions
while periodic boxes along the $y$ direction shift $\pm L_y \dot \gamma t$,
above and below the principal box, respectively.
Therefore, if a particle crosses $y=L_y/2$ to outside,
it enters the principal box again at $y=-L_y/2$ with $x$ shifted $-L_y \dot \gamma t$,
and $v_x$ shifted by $-L_y \dot \gamma$;
if the particle crosses $y=-L_y/2$ to outside,
it enters the principal box again at $y=L_y/2$ with $x$ shifted $L_y \dot \gamma t$,
and $v_x$ shifted by $L_y \dot \gamma$.
Furthermore, the $x$ and $z$ positions are always wrapped back into the principal box due to 
the periodic boundaries.
This is the so-called Lees-Edwards boundary condition~\cite{Evans2008},
which degenerates to the usual periodic box when $\dot\gamma=0$.

In practice, we utilize an implementation of the deforming triclinic box 
for the periodic shear flow in the LAMMPS package~\cite{Plimpton1995}.
In nonequilibrium MD, it is the so-called SLLOD dynamics for the canonical ensemble~\cite{Evans2008}.
The technical difference is that DPD has a built-in pairwise thermostat
while MD relies on other classical thermostats, such as the Nos\'{e}-Hoover thermostat.

\section{Results}
\label{section_results}

To compare the results of simulations with that of the theory,
we perform DPD simulations in a box of $[0, L_x] \times [-L_y/2, L_y/2] \times [0, L_z]$
so that the mean velocity $v_x=\dot\gamma y$ is consistent with the definition of the analytical solutions.
The domain is a cube with size $L_x=L_y=L_z=10$.
Input parameters of DPD are given in Section~3\ref{section_dpd_force}.
In the simulations, we define the fluctuating velocity under stationary shear flow as
\begin{eqnarray}
\delta v_{\mu}({\bf x}, t) = v_{\mu}({\bf x}, t) - \dot \gamma \delta_{\mu x}\delta_{\sigma y} x_{\sigma}.
\label{eq_fluctuating velcotiy particle}
\end{eqnarray}
The ACFs in k-space are calculated as 
\begin{eqnarray}
<\widehat{\delta u}_{\sigma}(-{\bf k},t) \widehat{\delta u}_{\sigma}({\bf k}, t+\tau)> = 
 \frac{1}{N_{s}} \sum^{N_{s}}_{s=1} 
 <\widehat{\delta u}_{\sigma}(-{\bf k},t) <\widehat{\delta u}_{\sigma}({\bf k}, t+\tau),
 \label{eq_acf_discrete}
\end{eqnarray}
where direction $\sigma=1$, $2$, and $3$, and
$N_s$ is the number of independent simulation runs.
The Fourier transform in space is defined as
\begin{eqnarray}
 \widetilde{\delta {\bf v}}({\bf k}, t) &= &
 \frac{1}{N_p} \sum^{N_p}_{j=1} \delta {\bf v}({\bf x}_j, t) e^{i{\bf k}(t) \cdot {\bf x}_j(t) }, \label{eq_fourier_transform_discrete}\\
 \widehat{\delta u}_{\sigma}({\bf k},t) &=& \widetilde{\delta {\bf v}}({\bf k}, t) \cdot {\bf e}^{\sigma} \label{eq_wave_vector_map}
\end{eqnarray}
where $j$ is particle index and $N_p$ is the number of particles in each simulation.
Note that fluctuating velocities are projected on unit vectors in the wave vector coordinate via Eq.~(\ref{eq_wave_vector_map})
after transformed in Eq.~(\ref{eq_fourier_transform_discrete}).

At equilibrium, there is no time origin,
therefore time averaging may be performed before 
ensemble averaging in Eq.~(\ref{eq_acf_discrete})
so that good statistics are obtained.
At nonequilibrium,
due to the time dependence of ${\bf k}(t)$,
basis vectors ${\bf e}^{\sigma}$ for wave vector coordinate is also time dependent~(see Appendix~\ref{appendix_hydrodynamic_matrix_and_modes}),
and hence it is generally much more expensive to reduce the statistical errors of ACFs.
Previously we have demonstrated that when the wave wave ${\bf k}_0=(0, 0, k_z \neq  0)$ is along the vorticity direction~\cite{Bian2016},
the two TACFs under shear flow degenerate to be isotropic just as that of equilibrium.
Therefore, in the following results, we shall consider $k_z=0$ and focus on the wave vectors
within the shear plane.

\subsection{Transversal autocorrelation functions: ${\bf k}_0=(2\pi/L_x, 0, 0)$}
\label{section_results_tacf_pzz}
If a wave vector along $x$ direction as ${\bf k}_0=(2\pi/L_x, 0, 0)$ is selected,
the dissipation of TACFs according to Eq. (\ref{eq_alpha}) is
\begin{equation}
  \alpha({\bf k}, \tau) = k^2_0 \tau + \frac{1}{3}\dot \gamma^2 k^2_x \tau^3. 
\end{equation} 
Therefore, it is expected that the dissipation resembles the equilibrium behavior $\sim \tau$ at short time
while it is dominated by the advection behavior $\sim \tau^3$ at long time.
The distinction of two time regimes can be observed for both TACFs as shown in Fig.~\ref{fig_tacfs_pzz}.
Equilibrium results with $\dot \gamma=0$ are also shown as reference;
with increasing shear rate $\dot\gamma$,
the dissipation rate is enhanced.
We note that the two TACFs have different intercepts with the $x$ axis,
as the extra term $k_0/k(\tau)$ for $C_{T_1}({\bf k}, \tau)$ in Eq. (\ref{eq_tacf1}) causes a further stronger decay.

For $\dot \gamma = 1.0$, the condition of $\dot\gamma \lesssim \nu k^2\approx 0.213$ is violated
and we can clearly see that $C_{T_1}({\bf k},\tau)$ of the simulations deviates from the theory of Eq. (\ref{eq_tacf1}).
For $\dot \gamma = 0.5$, although the condition of $\dot\gamma \lesssim \nu k^2$ is not strictly satisfied,
$\dot \gamma$ can still be treated as in the order of $k^2$ in the perturbation method.
Therefore, the TACFs for $\dot \gamma \le 0.5$ from the simulations agree with those from the theory very well.

 \begin{figure}
 \centering
  \includegraphics[width=0.49\textwidth]{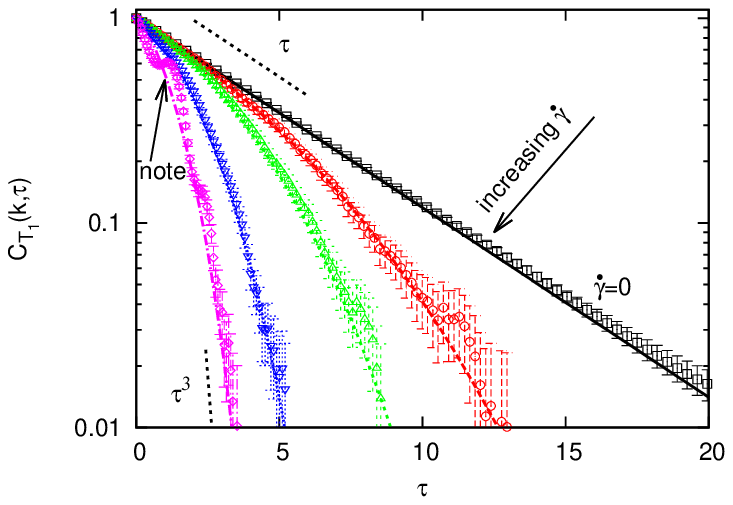}
  \includegraphics[width=0.49\textwidth]{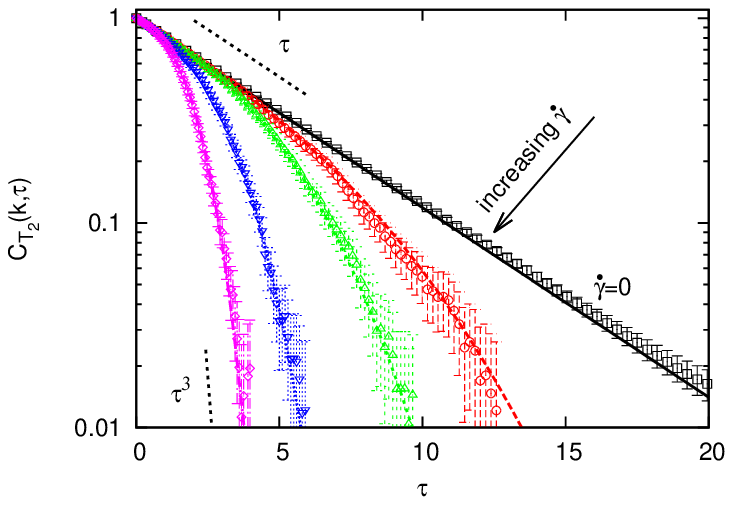}
  \caption{Transversal autocorrelation functions (TACFs) for ${\bf k}_0=(2\pi/L_x, 0, 0)$. 
  Left: $C_{T_1}({\bf k},t)$. Right : $C_{T_2}({\bf k},t)$.
  $\dot \gamma=1.0$, $0.5$, $0.2$, $0.1$ and $0$ (at equilibrium).
  Lines are from theory and symbols are from DPD simulations.
  Linear scale is along the $x$ axis and logarithmic scale is along the $y$ axis.}
  \label{fig_tacfs_pzz}
\end{figure}

\subsection{Transversal autocorrelation functions: ${\bf k}_0=(2\pi/L_x, \pm 2\pi/L_y, 0)$}
\label{section_results_tacf_ppnz}
If a wave vector is selected within the shear plane with both nonzero $x$ and $y$ components,
the dissipation rate according to Eq. (\ref{eq_alpha}) is
\begin{equation}
  \alpha({\bf k}, \tau) = k^2_0\tau-\dot \gamma k_x k_y \tau^2 + \frac{1}{3}\dot \gamma^2 k^2_x \tau^3. 
  \label{eq_alpha_ppz}
\end{equation} 
If $k_y>0$, for example, ${\bf k}_0=(2\pi/L_x, 2\pi/L_y, 0)$,
then the negative $\tau^2$ term competes with positive $\tau$ and $\tau^3$ terms.
Therefore, compared to the case of equilibrium, 
dissipation may decrease or increase at different time regimes.
This is well depicted for both $C_{T_1}({\bf k}, t)$ and $C_{T_2}({\bf k}, t)$ in Fig. \ref{fig_tacfs_ppz}.
It is noteworthy that the discrepancy between $C_{T_1}({\bf k}, t)$ of simulations and that of theory at $\dot \gamma=1$
is again apparent.

 \begin{figure}
 \centering
  \includegraphics[width=0.49\textwidth]{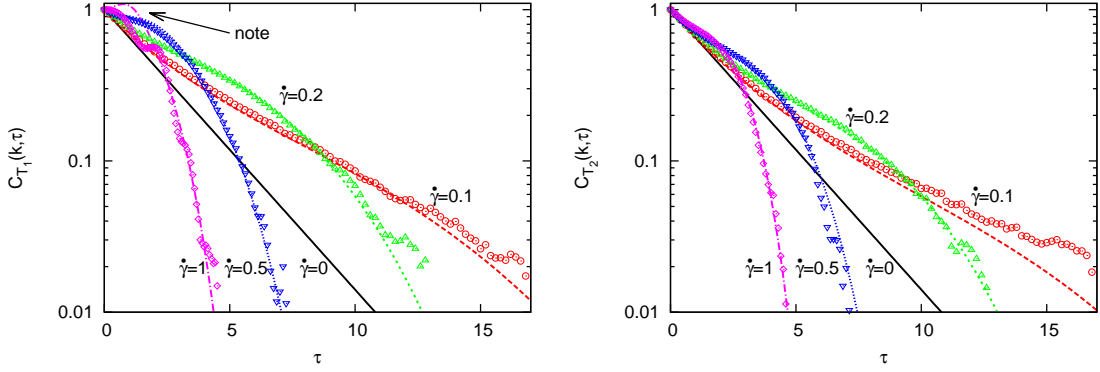}
  \includegraphics[width=0.49\textwidth]{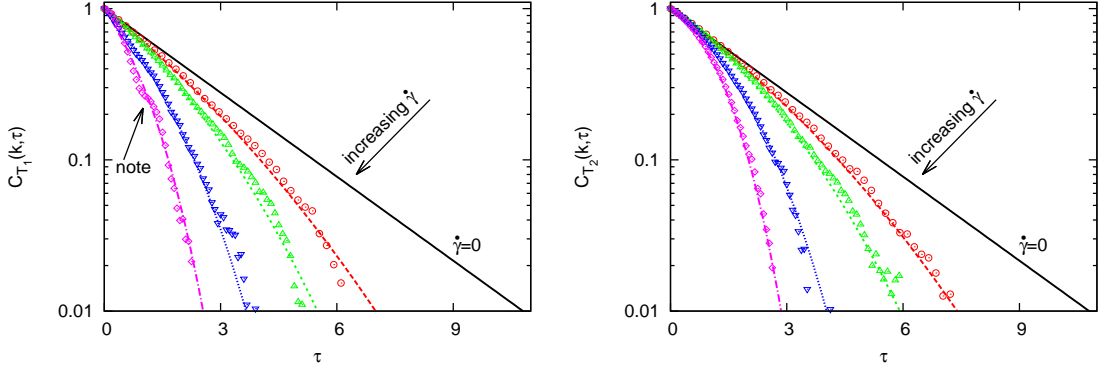}
  \caption{Transversal autocorrelation functions (TACFs) for ${\bf k}_0=(2\pi/L_x, 2\pi/L_y, 0)$. 
  Left: $C_{T_1}({\bf k},t)$. Right : $C_{T_2}({\bf k},t)$.
  $\dot \gamma=1.0$, $0.5$, $0.2$, and $0.1$.
  Lines are from theory and symbols are from DPD simulations.
  For $\dot\gamma=0$ at equilibrium, only the theory is plotted in solid lines as reference.
   Linear scale is along the $x$ axis and logarithmic scale is along the $y$ axis.}
   \label{fig_tacfs_ppz}
 \end{figure}

If a wave vector ${\bf k}_0=(2\pi/L_x, -2\pi/L_y, 0)$ is selected,
the three terms with different powers of $\tau$ in Eq. (\ref{eq_alpha_ppz}) are all positive.
Therefore, only enhancement of dissipation is expected compared to the case of equilibrium,
which is  confirmed in Fig. \ref{fig_tacfs_pnz}.
A slight discrepancy is noted for $C_{T_1}({\bf k}, \tau)$ between the simulations and the theory at $\dot\gamma=1$.

 \begin{figure}
 \centering
  \includegraphics[width=0.49\textwidth]{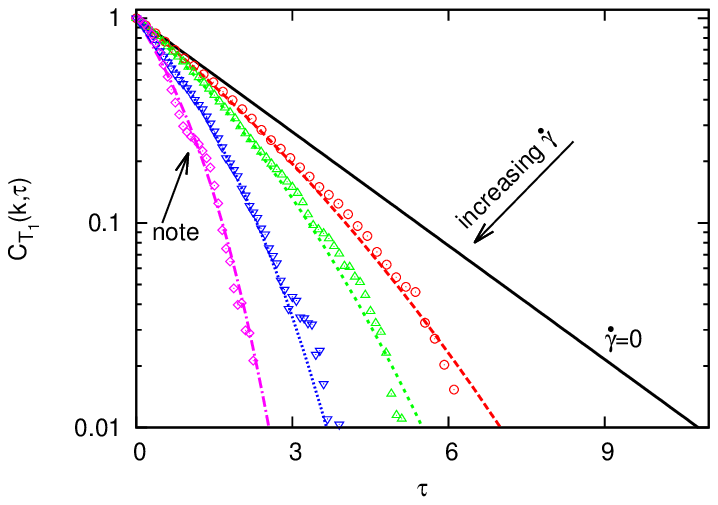}
  \includegraphics[width=0.49\textwidth]{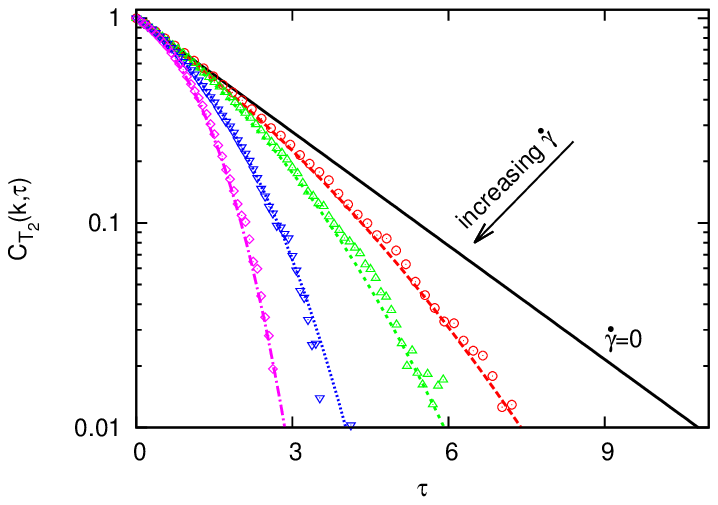}
  \caption{Transversal autocorrelation functions (TACFs) for ${\bf k}_0=(2\pi/L_x, -2\pi/L_y, 0)$.
  Left: $C_{T_1}({\bf k},t)$. Right : $C_{T_2}({\bf k},t)$.
  $\dot \gamma=1.0$, $0.5$, $0.2$, and $0.1$.
  Lines are from theory and symbols are from DPD simulations.
  For $\dot\gamma=0$ at equilibrium, only the theory is plotted in solid lines as reference.
  Linear scale is along the $x$ axis and logarithmic scale is along the $y$ axis.}
  \label{fig_tacfs_pnz}.
 \end{figure}

\subsection{Longitudinal autocorrelation functions: ${\bf k}_0=(2\pi/L_x, 0, 0)$ and ${\bf k}_0=(2\pi/L_x, \pm 2\pi/L_y, 0)$}
\label{section_results_lacf}
In this section, we evaluate the LACFs from both the DPD simulations and the theory. 
As indicated in Eq. (\ref{eq_lacf}), the LACF has both damping and oscillating elements.
The damping is primarily determined by the dissipation rate described by $\alpha({\bf k}, \tau)$ function given in Eq. (\ref{eq_alpha})
and much of its behavior has been seen in Sections \ref{section_results_tacf_pzz} and \ref{section_results_tacf_ppnz}.
Here we shall focus on the frequency of sound propagation,
which is determined by the $\beta ({\bf k}, \tau)$ function and it is repeated here
 \begin{equation}
   \beta({\bf k}, \tau) = \frac{1}{2\dot \gamma k_x} \left \{  k_yk_0-k_y(\tau)k(\tau)
-k^2_{\bot} In \left[ \frac{k_y(\tau)+k(\tau)}{k_y+k_0} \right]  \right \}.
 \end{equation}
When the wave vector is ${\bf k}_0=(2\pi/L_x, 0, 0)$ along $x$ direction,
the sound frequency is expected to increase monotonically with shear rate $\dot \gamma$,
as already indicated in Fig. \ref{fig_frequency}(a).
We further show the LACFs of both DPD simulations and the theory for this wave vector in Fig. \ref{fig_lacfs}(a).
We clearly observe an overall agreement between the results of DPD simulation and the theory,
and the sound frequency indeed increases monotonically with $\dot \gamma$.
A small discrepancy between the simulation and the theory is observed for $\dot \gamma=1$.

When the wave vector is ${\bf k}_0=(2\pi/L_x, 2\pi/L_y, 0)$, it is already suggested in Fig. \ref{fig_frequency}(b)
that the sound frequency may increase or decrease at different time regimes.
This is explicitly confirmed by the LACFs of both DPD simulations and the theory as shown in Fig. \ref{fig_lacfs}(b).
We may observe that simulations agree with the theory at all shear rates considered.

\begin{figure}
\centering
 \begin{subfigure}[b]{0.55\columnwidth}
  \includegraphics[width=\columnwidth]{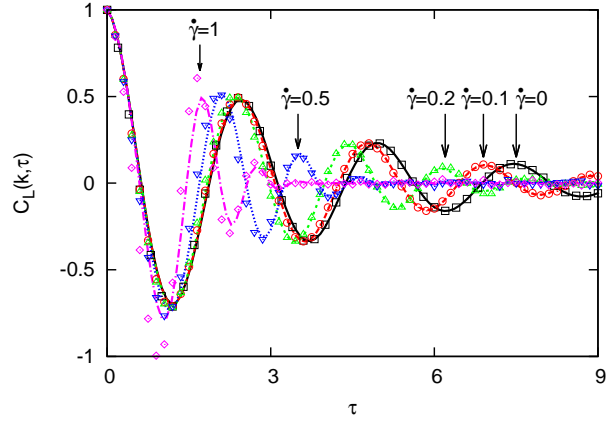}
  \caption{${\bf k}_0=(2\pi/L_x, 0, 0)$}
  \end{subfigure}
  \begin{subfigure}[b]{0.55\columnwidth}
  \includegraphics[width=\columnwidth]{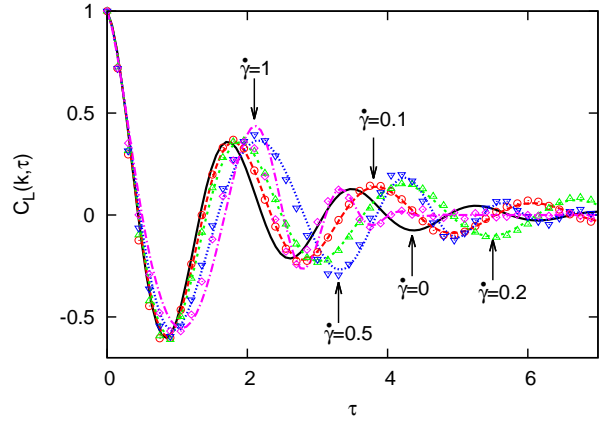} 
  \caption{${\bf k}_0=(2\pi/L_x, 2\pi/L_y, 0)$}
  \end{subfigure}  
  \begin{subfigure}[b]{0.55\columnwidth}
  \includegraphics[width=\columnwidth]{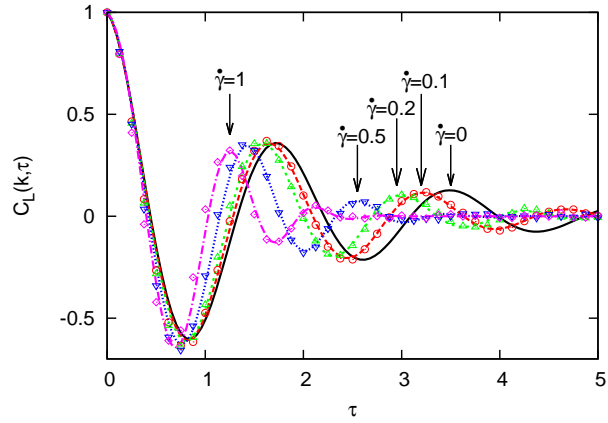} 
  \caption{${\bf k}_0=(2\pi/L_x, -2\pi/L_y, 0)$}
  \end{subfigure}    
  \caption{Longitudinal autocorrelation functions (LACFs) $C_L({\bf k}, t)$.
  $\dot \gamma=1.0$, $0.5$, $0.2$, and $0.1$.
  Lines are from theory and symbols are from DPD simulations.
  For $\dot\gamma=0$ at equilibrium, the theoretical line is also plotted as a reference.}
  \label{fig_lacfs}
\end{figure}
 
When the wave vector is ${\bf k}_0=(2\pi/L_x, -2\pi/L_y, 0)$, it is known from Fig. \ref{fig_frequency}(c) 
that sound frequency increases with $\dot \gamma$.
Again this can be observed in the LACFs in Fig. \ref{fig_lacfs}(c),
where the results of the simulations agree with the theory at all shear rates considered.
 
%\clearpage
 
\subsection{Small length scales at $\dot \gamma=1$}
\label{section_results_small_scales}
Here we revisit various ACFs at $\dot \gamma=1$,
where some discrepancies between the simulations and the theory 
have been observed for the largest length scale or smallest wave number considered.
For those cases, the condition $\dot \gamma \lesssim \nu k^2$ is not met.
Instead, here we focus on wave vectors ${\bf k}_0=(2n_w\pi/L_x, 0, 0)$ and ${\bf k}_0=(2n_w\pi/L_x, \pm 2n_w\pi/L_y, 0)$,
but with $n_w=2$ or $3$.
Therefore, the condition $\dot \gamma \lesssim \nu k^2$ becomes valid again for the smaller length scales and
we expect that the perturbation method in deriving the analytical solutions is accurate.
This can be confirmed by the TACFs $C_{T_1}({\bf k}, t)$ for different wave vectors with $n_w=2$ and $3$,
as shown in Fig. \ref{fig_tacfs_nw123}.
The same results of $n_w=1$ as in Fig. \ref{fig_tacfs_pzz}(a), \ref{fig_tacfs_ppz}(a) and \ref{fig_tacfs_pnz}(a)
are plotted again in Fig. \ref{fig_tacfs_nw123} for comparison.

 \begin{figure}\centering
 \begin{subfigure}[b]{0.55\columnwidth}
  \includegraphics[width=\columnwidth]{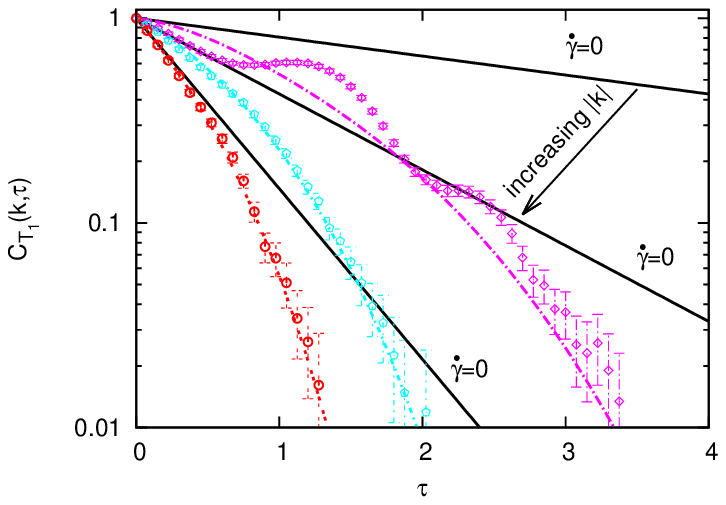}
  \caption{${\bf k}_0=(2n_w\pi/L_x, 0, 0)$}
  \end{subfigure}
  \begin{subfigure}[b]{0.55\columnwidth}
  \includegraphics[width=\columnwidth]{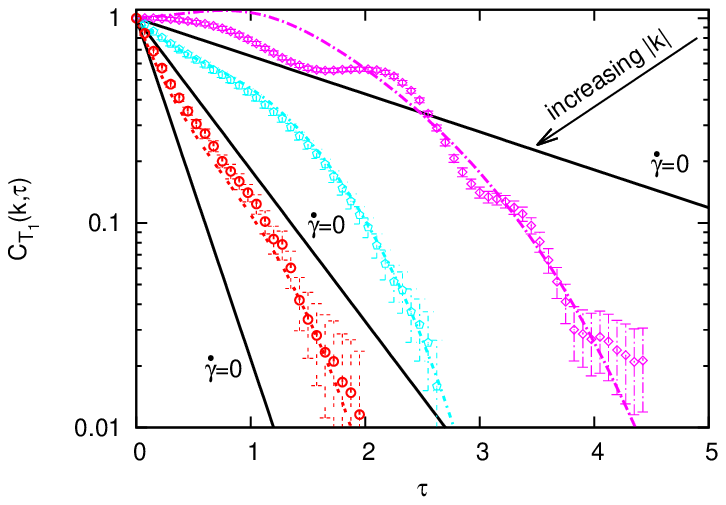} 
  \caption{${\bf k}_0=(2n_w\pi/L_x, 2n_w\pi/L_y, 0)$}
  \end{subfigure}  
  \begin{subfigure}[b]{0.55\columnwidth}
  \includegraphics[width=\columnwidth]{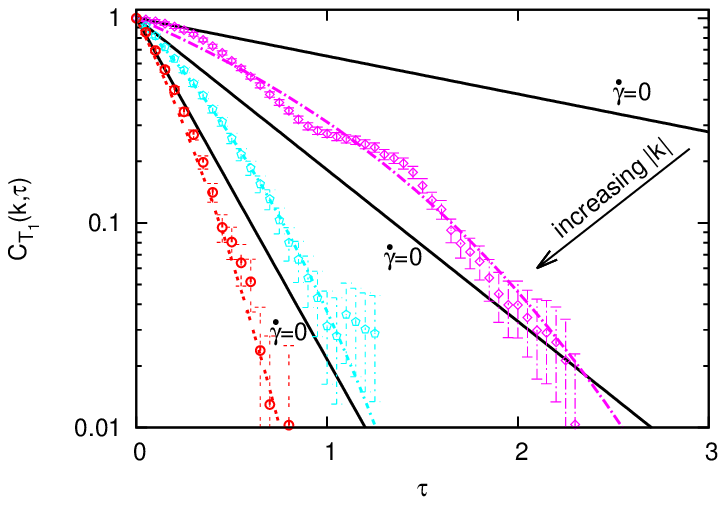} 
  \caption{${\bf k}_0=(2n_w\pi/L_x, -2n_w\pi/L_y, 0)$}
  \end{subfigure}    
  \caption{Transversal autocorrelation functions (TACFs) $C_{T_1}({\bf k}, t)$ at $\dot \gamma=1$:
  Wave number $n_w = 1, 2$, and $3$. 
  Lines are from theory and symbols are from DPD simulations.
  The theoretical results for $\dot \gamma=0$ at equilibrium are also plotted in solid lines for reference.
  Results of $n_w=1$ are the same as in Fig. \ref{fig_tacfs_pzz}(a), \ref{fig_tacfs_ppz}(a), and \ref{fig_tacfs_pnz}(a).}
  \label{fig_tacfs_nw123}
 \end{figure}
 
Similarly for the LACFs at $n_w=2$ and $3$, results of the simulations agree well with those of the theory,
since the condition $\dot \gamma \lesssim \nu k^2$ is valid, as shown in Fig. \ref{fig_lacfs_nw123}.
The same results of $n_w=1$ as in Fig. \ref{fig_lacfs} are shown again in Fig. \ref{fig_lacfs_nw123}
for comparison.

\begin{figure}\centering
 \begin{subfigure}[b]{0.55\columnwidth}
  \includegraphics[width=\columnwidth]{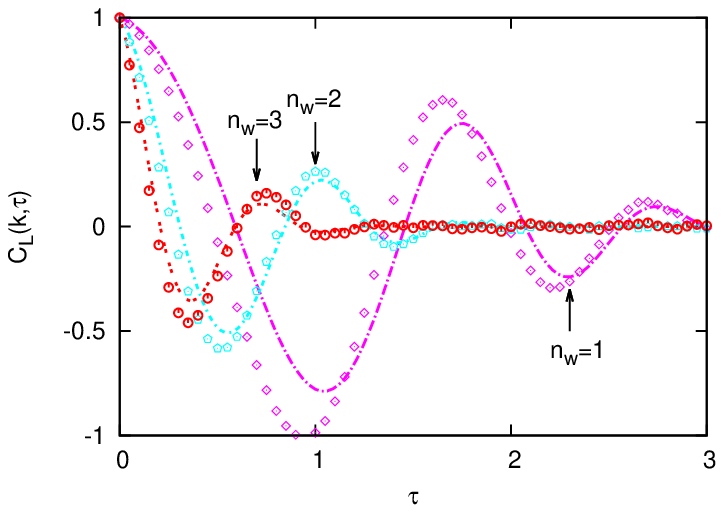}
  \caption{${\bf k}_0=(2n_w\pi/L_x, 0, 0)$}
  \end{subfigure}
  \begin{subfigure}[b]{0.55\columnwidth}
  \includegraphics[width=\columnwidth]{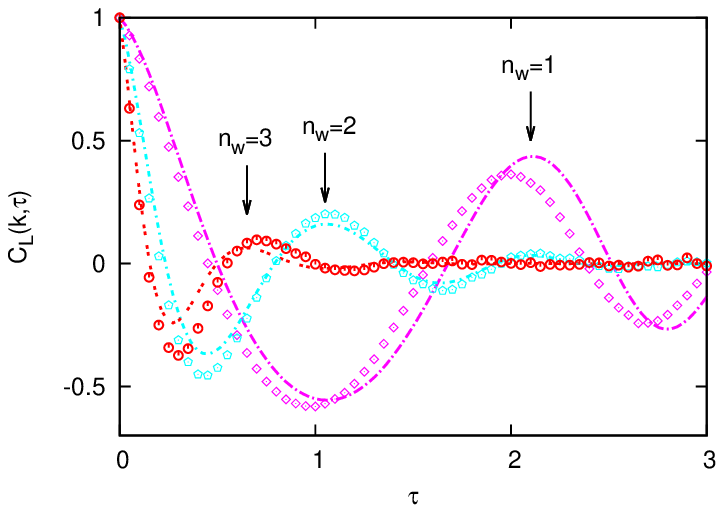} 
  \caption{${\bf k}_0=(2n_w\pi/L_x, 2n_w\pi/L_y, 0)$}
  \end{subfigure}  
  \begin{subfigure}[b]{0.55\columnwidth}
  \includegraphics[width=\columnwidth]{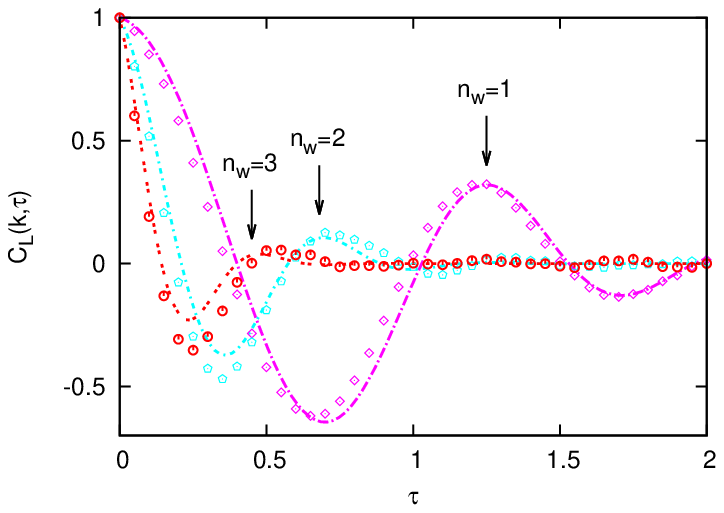} 
  \caption{${\bf k}_0=(2n_w\pi/L_x, -2n_w\pi/L_y, 0)$}
  \end{subfigure}    
  \caption{Longitudinal autocorrelation functions (LACFs) $C_{L}({\bf k}, t)$ at $\dot \gamma=1$:
  Wave number $n_w = 1, 2$, and $3$.
  Lines are from theory and symbols are from DPD simulations.
  Results of $n_w=1$ are the same as in Fig. \ref{fig_lacfs}.}
  \label{fig_lacfs_nw123}
 \end{figure}
 
It is noteworthy that at an even smaller scale ($n_w>3$),
another type of discrepancy between the simulations and the theory emerges.
This is due to the fact that 
the classical fluctuating hydrodynamics may not describe well
the collective behavior of the DPD simulations at this small scales~\cite{Ripoll2001, Azarnykh2016}.
In this case, solutions of the generalized fluctuating hydrodynamics may be required~\cite{Boon1991},
which is beyond the scope of the current work.

\section{Discussion and summary}
\label{section_summary}

We studied the autocorrelations (ACFs) of hydrodynamic fluctuations in k-space for an isothermal fluid 
under shear flow, which is driven by the Lees-Edwards periodic boundary condition.
We compared results of the ACFs from the dissipative particle dynamics (DPD) simulations with the theoretical approximations.
DPD is a particle-based mesoscopic method with three pairwise forces between neighboring particles.
After specifying the conservative, dissipative and random forces, 
DPD is an effective simulator for the compressible fluctuating hydrodynamics.
Under the assumption of a local thermodynamic equilibrium,
the dissipative and random forces act as an intrinsic thermostat 
and they satisfy the fluctuation-dissipation theorem.
In principle, the DPD method is valid for an arbitrarily large shear rate $\dot \gamma$,
so are its ACFs of fluctuations,
as long as local thermodynamic equilibrium is valid.

In order to derive the analytical solutions for the various ACFs,
a perturbation method is adopted to expand the eigenvalue and eigenvector as a power series of $k$,
where the parameter $k$ is the inverse of wave-length
and it is very small in the hydrodynamic limit.
In such regime, $\nu k^2 \ll c_Tk$ holds so that
the rate of hydrodynamic dissipation is much smaller than the rate of sound propagation.
Here $\nu$ and $c_T$ are the kinematic viscosity and isothermal sound speed, respectively.
In the perturbation method, it is assumed that $\dot \gamma \lesssim \nu k^2$.
Therefore, terms of $\dot \gamma$ are treated as in the  order of $k^2$.
With such an assumption, the generalized eigenvalue problem was solved approximately and
thereafter the longitudinal and the two transversal ACFs were constructed
as in Eqs. (\ref{eq_lacf}), (\ref{eq_tacf1}), and (\ref{eq_tacf2}).

When the condition of $\dot \gamma \lesssim \nu k^2$ is met,
that is, Reynolds number $Re=\dot \gamma L^2/ \nu \lesssim 4\pi^2$,
various ACFs of the analytical solutions are accurate
and agree very well with those of DPD simulations.
We observed that the two transversal ACFs in shear flow are no longer identical as in equilibrium
and the dissipation rate is time dependent as well.
Furthermore, depending on the individual wave vector ${\bf k}$,
enhancement or attenuation of sound frequency may take place
and we observed the Doppler effects.
Given the increasing efforts on hybrid modeling of fluid flow,
where usually a stochastic microscopic/mesoscopic solver
is concurrently coupled with a continuum (Landau-Lifshitz-)Navier-Stokes solver~\cite{Delgado-Buscalioni2012, Bian2016},
the agreement between analytical and computational approaches
on the results is meaningful:
the temporal correlations of fluctuating variables
provide a fundamental benchmark for any proposed coupling algorithm.

When $\nu k^2 < \dot \gamma < c_T k$, some discrepancies 
between the ACFs of the analytical solutions and those of the DPD simulations are observed.
In this regime, besides the coupling between the advection and fluctuating variables,
extra couplings between different fluctuating variables at equal time are expected to be significant.
In our DPD simulations these extra couplings affect significantly both $C_{T_1}$ and $C_L$,
but have a negligible effect on $C_{T_2}$.
In this regime of shear rates,
the perturbation theory is inaccurate and should be modified to treat the terms of $\dot \gamma$ as in the order of $k$
instead of $k^2$.
Furthermore, the contributions from the stochastic stress
on the temporal correlations are also expected to emerge.
However, such modifications on the theory are nontrivial to accomplish and are subjects of our future research.
Nevertheless, the DPD simulations are in principle valid and
provide a few guidelines on how to improve the theory.
In this regime, comparison with results from other numerical methods
such as the finite volume method with thermal fluctuations~\cite{DeFabritiis2007, Chaudhri2014} would also be helpful
to confirm the results presented.

When the shear rate becomes even larger such as $ \dot \gamma > c_Tk$,
we obtain a nonequilibrium state far from equilibrium
and additional couplings between different $k$ are expected~(results are not shown).
The local thermodynamic equilibrium is produced and enforced by the frequent collisions between particles
and the frequency is measured by the sound speed $c_T$ divided by the length scale, that is $c_Tk$.
Therefore, under such strong shear flow the local thermodynamic equilibrium can no longer be restored in the length scale of $1/k$.
It would also be interesting to evaluate the probability of violating the Second law in such a DPD system.

\section*{Acknowledgments}
This work was supported by the Computational Mathematics Program
within the Department of Energy office of Advanced Scientific Computing Research as part of the
Collaboratory on Mathematics for Mesoscopic Modeling of Materials (CM4) 
and also supported by the ARO grant W911NF-14-1-0425.
Part of this research was conducted using computational resources 
and services at the Center for Computation and Visualization, Brown University.
An award of computer time was provided by the Innovative and Novel Computational Impact on Theory and Experiment (INCITE) program. 
This research used resources of the Argonne Leadership Computing Facility, 
which is a DOE Office of Science User Facility supported under contract DE-AC02-06CH11357. 
This research also used resources of the Oak Ridge Leadership Computing Facility, 
which is a DOE Office of Science User Facility supported under Contract DE-AC05-00OR22725.
X. B. acknowledges discussions with Dr. Fanhai Zeng on the generalized eigenvalue problem.

%\disclaimer{Insert disclaimer text here.}

\appendix

\section{Linearization around uniform shear flow}
\label{appendix_linearization}

The fluctuations on the state variables are defined as
\begin{eqnarray}
{\bf z}({\bf x},t)=[\delta \rho({\bf x},t), \delta{\bf v}({\bf x},t)],
\end{eqnarray}
with
\begin{eqnarray}
 \delta \rho     &=& \rho  -\rho_0        =  \rho - <\rho>, \\
  \delta {\bf v} &=& {\bf v}- {\bf v}_0   = {\bf v} - <{\bf  v}>,
\end{eqnarray}
where the macroscopic state variables are the averaged quantities denoted by ``$< \quad >$''.
Therefore, the fluctuating hydrodynamic equations can be linearized as
\begin{eqnarray}
 \left( \frac{\partial}{\partial t} + \dot \gamma_{\mu\sigma} x_{\sigma} \frac{\partial}{\partial x_{\mu}}\right) \delta \rho 
 + \rho_0 \nabla \cdot \delta{\bf v} &=& 0, \label{eq_continuity_Eulerian_linear} \\
 \left( \frac{\partial}{\partial t} + \dot \gamma_{\mu\sigma} x_{\sigma} \frac{\partial}{\partial x_{\mu}}\right) \delta v_{\mu} 
 + \dot \gamma_{\mu\sigma} \delta v_{\sigma} + \frac{c^2_T}{\rho_0}\frac{\partial}{\partial x_{\mu}}\delta \rho  \nonumber \\
 - \nu \nabla^2 \delta v_{\mu} 
 - \left( \kappa + \frac{\nu}{3} \right)\frac{\partial}{\partial x_{\mu}} \nabla  \cdot \delta {\bf v} &=& 
 \frac{1}{\rho_0} \frac{\partial}{\partial x_{\mu}} \Pi^R_{\mu\sigma}, \label{eq_momentum_Eulerian_linear} 
\end{eqnarray}
where second-order terms in fluctuations are neglected.
Kinematic viscosities are defined as $\nu=\eta/\rho_0$ and $\kappa=\zeta/\rho_0$.
Furthermore, the equation of state is assumed to have the property of $\delta p = c^2_T \delta \rho$,
where $c_T$ is the isothermal sound speed.

\section{Hydrodynamic matrix}
\label{appendix_hydrodynamic_matrix_and_modes}

It proves to be convenient to solve such a linearized hydrodynamic equations of Eqs.~(\ref{eq_continuity_Eulerian_linear}) and (\ref{eq_momentum_Eulerian_linear})
in Fourier space~\cite{Kadanoff1963, Boon1991, Hansen2013}.
Hence, we define the fluctuations in k-space as the spatial Fourier transform of the fluctuations
\begin{eqnarray}
 \widetilde z({\bf k}, t) = \int z({\bf x}, t) e^{i {\bf k}\cdot {\bf x}} d{\bf x}.
\end{eqnarray}
Suppose that the wave vector is defined as ${\bf k}=k_x{\bf e}_x + k_y{\bf e}_y + k_z{\bf e}_z= (k_x, k_y, k_z)$,
where ${\bf e}_x$, ${\bf e}_y$ and ${\bf e}_z$ are three basis vectors in the fixed Cartesian coordinate.
The formulation of the problem is very succinct, 
if the transformed velocity is decomposed into 
one component parallel to ${\bf k}$ and the other two components perpendicular to ${\bf k}$~\cite{Lutsko1985, Otsuki2009}.
Therefore, we define another three orthonormal vectors ${\bf e}^1$, ${\bf e}^2$ and ${\bf e}^3$
in such way that ${\bf e}^1$ is along ${\bf k}$
and ${\bf e}^{2,3}$ are perpendicular to ${\bf k}$,
\begin{eqnarray}
 {\bf e}^1 &=& {\bf k}/k, \label{eq_wave_unit_vector1} \\
 {\bf e}^2 &=& \left[{\bf e}_y - ({\bf e}^1 \cdot {\bf e}_y) {\bf e}^1 \right]/k_{\bot}, \label{eq_wave_unit_vector2} \\
 {\bf e}^3 &=& {\bf e}^1 \times {\bf e}^2.
 \label{eq_wave_unit_vector3}
\end{eqnarray}
Here ${\bf e}_y$ is taken as reference to define the first transversal direction ${\bf e}^2$
so that ${\bf e}^2$ is in the same plane with ${\bf e}^1$ and ${\bf e}_y$,
and ${\bf e}^2$ is perpendicular to ${\bf e}^1$;
moreover, ${\bf e}^3$ is perpendicular to both ${\bf e}^1$ and ${\bf e}^2$.
Also $k=|{\bf k}|$, $k_{\bot}=(k^2-k^2_y)^{1/2}/k$.
Note that superscript $1$, $2$, or $3$ refers to the pairwise orthonormal basis vectors in the wave vector coordinate
or oblique coordinate,
which are expressed in the fixed Cartesian coordinate.

We further define a vector of fluctuating variables in k-space as
\begin{eqnarray}
 \widehat {\bf z}({\bf k}, t) &=&  \left[ 
 \widehat{\delta {\rho}}({\bf k},t),
 \widehat{\delta u}_1({\bf k},t), 
 \widehat{\delta u}_2({\bf k},t), 
 \widehat{\delta u}_3({\bf k},t) \right],
\end{eqnarray}
where each element is related to the Fourier-transformed variable as
\begin{eqnarray} 
 \widehat{\delta \rho}({\bf k}, t) &=& c_T\widetilde{\delta {\rho}}({\bf k},t)/\rho_0, \\
 \widehat{\delta u}_{\mu}({\bf k}, t) &=& \widetilde{\delta {\bf v}}({\bf k},t) \cdot {\bf e}^{\mu}.\label{eq_velocity_fourier_component}
\end{eqnarray}
The fluctuating velocity $\widetilde{\delta {\bf v}}({\bf k},t)$ in k-space
is the spatial Fourier transform of the fluctuating velocity $\delta {\bf v}({\bf x},t)$ in real space.
The latter is the peculiar or fluctuating velocity around the macroscopic velocity field defined as
\begin{eqnarray}
\delta v_{\mu}({\bf x}, t) = v_{\mu}({\bf x}, t) - \dot \gamma \delta_{\mu x}\delta_{\sigma y} x_{\sigma}.
\label{eq_fluctuating_velocity}
\end{eqnarray}
It is also simple to see that $\widehat{\delta {\bf u}}({\bf k},t)$ is obtained
by projecting $\widetilde{\delta {\bf v}}({\bf k},t)$ onto the basis directions of the wave vector.
Hence, $\widehat{\delta u}_1({\bf k}, t)$ is called the longitudinal fluctuating component,
while $\widehat{\delta u}_2({\bf k}, t)$ and $\widehat{\delta u}_3({\bf k}, t)$
are the first and second transversal fluctuating components, respectively.
Finally, the hydrodynamic equations in k-space are summarized in a compact form as
\begin{eqnarray}
 \left( \frac{\partial}{\partial t} - \dot \gamma_{\mu\sigma} k_{\mu} \frac{\partial}{\partial k_{\sigma}} \right) \widehat z_{\epsilon}({\bf k}, t)
 + {\mathcal L}_{\epsilon \iota}({\bf k}, \dot \gamma, t) \widehat z_{\iota}({\bf k}, t) = \widehat {\bf R}_{\epsilon}({\bf k}, t),
 \label{eq_hydrodynamic_equation}
\end{eqnarray}
%where the detailed expressions for the linear operator ${\mathcal L}$ and 
%the random term $\widehat {\bf R}$ are given in Appendix \ref{appendix_matrix}.
where the matrix ${\mathcal L}$ is defined as
\begin{eqnarray}
 {\mathcal L}=-ik{\mathcal A}+k^2{\mathcal B}+\dot\gamma{\mathcal C},
 \label{eq_hydrodynamic_matrix}
\end{eqnarray}
with
\begin{eqnarray}
 {\mathcal A} = \left( 
 \begin{array}{cccc}
  0   & c_T & 0 & 0\\
  c_T & 0   & 0 & 0\\
  0   & 0   & 0 & 0\\
  0   & 0   & 0 & 0
 \end{array}
 \right),
\end{eqnarray}
\begin{eqnarray}
 {\mathcal B} = \left( 
 \begin{array}{cccc}
  0   & 0   & 0 & 0\\
  0   & \nu_L & 0 & 0\\
  0   & 0   & \nu & 0\\
  0   & 0   & 0 & \nu
 \end{array}
 \right),
\end{eqnarray}
and
\begin{eqnarray}
 {\mathcal C} = \left( 
 \begin{array}{cccc}
  0   & 0   & 0 & 0\\
  0   & \Phi_{11} & \Phi_{12} & \Phi_{13}\\
  0   & \Phi_{21}  & \Phi_{22} & \Phi_{23}\\
  0   & \Phi_{31}  & \Phi_{32} & \Phi_{33}
 \end{array}
 \right).
\end{eqnarray}
The longitudinal viscosity is defined as $\nu_L=4\nu/3 + \kappa=(4\eta/3+\zeta)/\rho_0$. 
$\Phi_{\mu\sigma}$ is defined by
\begin{eqnarray}
\dot\gamma \Phi_{\mu\sigma}= 
{\bf e}^{\mu}_{\epsilon}({\bf k}) \dot \gamma_{\epsilon\iota} {\bf e}^{\sigma}_{\iota}({\bf k})
-  \dot \gamma_{\epsilon\iota} k_{\epsilon} {\bf e}^{\mu}_{\chi} ({\bf k}) \frac{\partial}{\partial k_{\iota}}{\bf e}^{\sigma}_{\chi}({\bf k}).
\end{eqnarray}
Given the definitions of the basis vectors from Eqs.~(\ref{eq_wave_unit_vector1}), 
(\ref{eq_wave_unit_vector2}), and (\ref{eq_wave_unit_vector3}) in oblique coordinate,
the elements of ${\mathcal C}$ read
\begin{eqnarray}
 \Phi_{11}&=&-\Phi_{22}=k_xk_y/k^2,\\
 \Phi_{12}&=&-k_x/k_{\perp},\\
 \Phi_{31}&=&-k_yk_z/kk_{\perp},\\
 \Phi_{32}&=&-k_z/k, \\
 \Phi_{ij}&=0&, \quad all \quad others.
\end{eqnarray}
The linear mode coupling is represented by the derivatives with respect to ${\bf k}$.

Finally, the stochastic terms are 
\begin{eqnarray}
 {\widehat R}_{\epsilon}({\bf k}, t) &=& \widetilde{R}_{\epsilon}({\bf k}, t) -  <\widetilde{R}_{\epsilon}({\bf k}, t)>,\\
 \widetilde{R}_{1}({\bf k}, t)     &=& 0,\\
 \widetilde{R}_{2}({\bf k}, t)   &=& {\bf e}^1_{\mu}({\bf k})ik_{\sigma} \widetilde\Pi^R_{\mu\sigma}({\bf k},t)/\rho_0,\\
 \widetilde{R}_{3}({\bf k}, t)   &=& {\bf e}^2_{\mu}({\bf k})ik_{\sigma} \widetilde\Pi^R_{\mu\sigma}({\bf k},t)/\rho_0,\\ 
 \widetilde{R}_{4}({\bf k}, t)   &=& {\bf e}^3_{\mu}({\bf k})ik_{\sigma} \widetilde\Pi^R_{\mu\sigma}({\bf k},t)/\rho_0.\\ 
\end{eqnarray}

The general solution to the linearized hydrodynamic equation (Eq.~\ref{eq_hydrodynamic_equation}) 
can be determined from the nonlinear eigenvalue problem,
\begin{eqnarray}
 \left( -\dot \gamma_{\mu\sigma}k_{\mu}\frac{\partial}{\partial k_{\sigma}} 
 + \mathcal L\right)\xi^{(\mu)} = \lambda_{\mu} \xi^{(\mu)}. \label{eq_eigenvalue_equation}
\end{eqnarray}
The left eigenvectors $\chi^{(\mu)}$ are defined as
\begin{eqnarray}
 \left(\chi^{(\mu)}, \xi^{(\sigma)} \right) = 
 \bar \chi^{(\mu)}_{\epsilon} \xi^{(\sigma)}_{\epsilon} = \delta_{\mu\sigma},\label{eq_eigenvector_left_right_biorthogonal}
\end{eqnarray}
where $\bar \chi^{(\mu)}_{\epsilon}$ means conjugate of $\chi^{(\mu)}_{\epsilon}$,
and $\delta_{\mu\sigma}$ is the Kronecker $\delta$

The eigenvalues $\lambda_{\mu}$ and eigenvectors $\xi^{(\mu)}$ can be
obtained via the perturbation theory by the expansion of $k$ 
to second and first order, respectively 
\begin{eqnarray}
 \lambda_{\mu} &=& k\lambda_{\mu,1}+k^2\lambda_{\mu,2} + \ldots  \quad , \label{eq_eigenvalue_expansion}\\
 \xi^{(\mu)} &=& \xi^{(\mu)}_0 + k \xi^{(\mu)}_1 + \ldots  \quad.\label{eq_eigenvector_expansion}
\end{eqnarray}
Inserting Eqs.~(\ref{eq_eigenvalue_expansion}) and (\ref{eq_eigenvector_expansion}) into 
Eq. (\ref{eq_hydrodynamic_equation}) and 
making use of the explicit form of $\mathcal L$ in Eq.~(\ref{eq_hydrodynamic_matrix}),
we have the first-order perturbation theory involving ${\mathcal A}$
and second-order perturbation theory involving ${\mathcal A}$, ${\mathcal B}$, and ${\mathcal C}$ all together.
Assuming $\dot \gamma \lesssim \nu k^2 \ll c_Tk$ 
and treating $\dot \gamma$ as in the order of $k^2$,
the two equations from the perturbation read~\cite{Lutsko1985, Lutsko1986a}
\begin{eqnarray}
 \left(-i{\mathcal A} - \lambda_{\mu,1} {\mathcal I}\right) \xi^{(\mu)}_0&=&0, \label{eq_eigenvalue_equation1}\\
 \left(-i{\mathcal A} - \lambda_{\mu,1} {\mathcal I}\right) \xi^{(\mu)}_1&=& \label{eq_eigenvalue_equation2}
 \left(\lambda_{\mu,2} {\mathcal I} - {\mathcal B}
 - \dot \gamma k^{-2} {\mathcal C} 
 + k^2 \dot \gamma_{\mu \sigma} k_{\mu} 
 \frac{\partial}{\partial k_{\sigma}}\right)\xi^{(\mu)}_0.
\end{eqnarray}
From Eq.~(\ref{eq_eigenvalue_equation1}), we find the first set of eigenvalues as follows
\begin{eqnarray}
 \lambda_{1,1} &=& -ic_T, \\
 \lambda_{2,1} &=& +ic_T, \\
 \lambda_{3,1} &=& \lambda_{4,1} = 0.
\end{eqnarray}
From Eq. (\ref{eq_eigenvalue_equation2}), we find the second set of eigenvalues as follows
\begin{eqnarray}
 \lambda_{1,2} &=& 
 =  \Gamma_T + \frac{1}{2} \dot \gamma k^{-2} \Phi_{11}
 =  \Gamma_T + \frac{1}{2} \dot \gamma k_xk_y/k^4,\\
 \lambda_{2,2} &=& \lambda_{1,1} \\
 \lambda_{3,2} &=& \nu + \dot \gamma k^{-2} \Phi_{22} = \nu - \dot \gamma k_xk_y/k^4, \\
 \lambda_{4,2} &=& \nu,
\end{eqnarray}
where sound attenuation coefficient $\Gamma_T=\nu_L/2=(2\eta/3+\zeta/2)/\rho_0$.

Inserting the expressions for $\lambda_{\mu,1}$ and $\lambda_{\mu,2}$ back into Eq. (\ref{eq_eigenvalue_expansion}),
finally the eigenvalues are summarized as
\begin{eqnarray}
 \lambda_1 &=& -ic_Tk +  \Gamma_T k^2 + \frac{1}{2} \dot\gamma k_xk_y/k^2,\\
 \lambda_2 &=& +ic_Tk +  \Gamma_T k^2 + \frac{1}{2} \dot\gamma k_xk_y/k^2,\\
 \lambda_3 &=& \nu k^2-\dot\gamma k_x k_y/k^2,\\
 \lambda_4 &=& \nu k^2.
\end{eqnarray}
Furthermore, the corresponding right eigenvectors are
 \begin{eqnarray}
 \xi^{(1)} &=& \frac{1}{\sqrt{2}}(1,1,0,0)^T,\\
 \xi^{(2)} &=& \frac{1}{\sqrt{2}}(0,0,0,1)^T,\\
 \xi^{(3)} &=& \left(0,0,1,M\right)^T,\\
 \xi^{(4)} &=& \left(0, 0, 0, 1 \right)^T,
\end{eqnarray}
where 
\begin{eqnarray}
 M({\bf k}) &=& \frac{kk_z}{k_xk_{\perp}}\arctan(\frac{k_y}{k_{\perp}}),\\
 k_{\perp}  &=& k^2-k^2_y.
\end{eqnarray}
The corresponding left eigenvectors satisfying Eq. (\ref{eq_eigenvector_left_right_biorthogonal}) are
\begin{eqnarray}
 \chi^{(1)} &=& \frac{1}{\sqrt{2}}(1,1,0,0)^T,\\
 \chi^{(2)} &=& \frac{1}{\sqrt{2}}(0,0,0,1)^T,\\
 \chi^{(3)} &=& \left(0,0,1,0\right)^T,\\
 \chi^{(4)} &=& \left(0, 0, -M, 1 \right)^T.
\end{eqnarray}

Given the eigenvalues and eigenvectors $\lambda_{\mu}$, $\xi^{\mu}$ and $\chi^{\mu}$, 
the time evolution of fluctuating variables is obtained as
\begin{eqnarray}
 \widehat z_{\epsilon}({\bf k}(t), t) = \sum^4_{\iota=1} G_{\epsilon \iota} ({\bf k}(t),t) \widehat z_{\iota}({\bf k}, 0)
 + \int^t_0 \sum^4_{\iota=1} G_{\epsilon \iota} ({\bf k}(t-u),t-u) \widehat R_{\iota}({\bf k}, u) du,
\end{eqnarray}
with a propagator defined as
\begin{eqnarray}
 G_{\epsilon \iota} ({\bf k}(t),t) = \sum^4_{\mu=1} \xi^{(\mu)}_{\epsilon}({\bf k}(t)) \chi^{(\mu)}_{\iota}({\bf k}) e^{-\int^t_0 d\tau \lambda_{\mu}({\bf k}(\tau))}.
 \label{eq_propagator}
\end{eqnarray}
To account for the advection, the wave vector is time dependent ${\bf k}(t)=(k_x, k_y-\dot\gamma t k_x, k_z)$.
Since the random terms do not contribute in this shear rate regime~\cite{Lutsko1985, Varghese2015},
the correlations of the fluctuating variables are readily expressed as
\begin{eqnarray}
C_{\epsilon \iota}({\bf k}, -{\bf k}, t)= <\widehat z_{\epsilon}({\bf k}(t), t) \widehat z_{\iota} (-{\bf k}, 0) > 
= \sum^4_{\mu=1} G_{\epsilon \mu}({\bf k}(t),t) <\widehat z_{\mu}({\bf k}, 0) \widehat z_{\iota}(-{\bf k}, 0) >.
\end{eqnarray}
Given the values of $\lambda_{\mu}$, $\xi^{\mu}$ and $\chi^{\mu}$,,
we find that $G_{\epsilon \iota}=0$ for $\epsilon \neq \iota$
and $G_{\epsilon \iota} \neq 0$ for $\epsilon = \iota$.

%%%%%%%%%% Insert bibliography here %%%%%%%%%%%%%%
%\bibliographystyle{unsrtnat}
\bibliographystyle{unsrt}
%\bibliographystyle{rspublicnat}
%\bibliographystyle{vancouver}
%\bibliographystyle{abbrvnat}
% \bibliography{../../../bibtex/bian_bibtex_jabref}
% \bibliography{/media/bianx/data1/LRZ_sync_and_share/publication/bibtex/bian_bibtex_jabref}

\end{document}